\documentstyle[12pt]{article}

\newcommand{\beq}{\begin{equation}}
\newcommand{\eeq}{\end{equation}}
\newcommand{\bq}{\begin{quotation}}
\newcommand{\eq}{\end{quotation}}
\newcommand{\bc}{\begin{center}}
\newcommand{\ec}{\end{center}}

\newcommand{\lr}[2]{\langle #1,#2 \rangle}

\newcommand{\lxixir} {{\lr{x_i}{x_i}}}
\newcommand{\lyiyir} {{\lr{y_i}{y_i}}}
\newcommand{\lxiyir} {{\lr{x_i}{y_i}}}
\newcommand{\lxiPir} {{\lr{x_i}{P_i}}}

\newcommand{\lUiyir} {{\lr{U_i}{y_i}}}
\newcommand{\lyiUir} {{\lr{y_i}{U_i}}}

\newcommand{\lPiPir} {{\lr{P_i}{P_i}}}
\newcommand{\lPiVir} {{\lr{P_i}{V_i}}}
\newcommand{\lPiyir} {{\lr{P_i}{y_i}}}
\newcommand{\lviyir} {{\lr{v_i}{y_i}}}
\newcommand{\lviUir} {{\lr{v_i}{U_i}}}
\newcommand{\lPUr} {{\lr{P}{U}}}

\newcommand{\lUUr} {{\lr{U}{U}}}

\newcommand{\lUVr} {{\lr{U}{V}}}
\newcommand{\lUiVir} {{\lr{U_i}{V_i}}}
\newcommand{\lUivir} {{\lr{U_i}{v_i}}}

\newcommand{\lVixir} {{\lr{V_i}{x_i}}}

\newcommand{\tr}   {\mbox{tr}}
\newcommand{\half} {\frac{1}{2}}
\newcommand{\r}    {\rightarrow}
\newcommand{\laB}{\lambda_B}
\newcommand{\PPsi}{{\bf X}} 
\newcommand{\dP}{{\delta \bf X}} 
\newcommand{\plaq}{{\Box}}

\newcommand{\simleq}{{\leq}}

\begin{document}

.

\vspace{4.0 cm}

\begin{center}

{\bf\Large {Chaos and Scaling in Classical Non-Abelian Gauge Fields}}

\vspace*{1.1ex}

{\bf \large H.B. Nielsen$^1$, H.H. Rugh$^2$ and S.E. Rugh$^3$ } \\

{\vspace*{1.0ex}
       $^1$ The Niels Bohr Institute, Blegdamsvej 17, 2100
       K\o benhavn \O, Denmark  \\ [1.2ex]
       $^2$ Department of Mathematics, University of Warwick,
        Coventry, CV4 7AL, England   \\ [1.2ex]
       $^3$ Theoretical Division, T-6, MS B 288, University of
        California, Los Alamos National Laboratory, Los Alamos,
         New Mexico, NM 87545, U.S.A. 
\\ [1.2ex]

}

\end{center}

\begin{abstract}
\noindent
Without an ultraviolet cut-off, the time evolution of 
the classical Yang-Mills equations give rise to a never ending cascading of
the modes towards the ultraviolet, and ergodic measures and 
dynamical averages, such as the
spectrum of characteristic
Lyapunov exponents (measures of temporal chaos) or spatial
correlation functions, are ill defined. 
A lattice regularization (in space) 
provides an ultraviolet cut-off of the classical Yang-Mills theory,
giving a possibility for the existence of ergodic measures and dynamical
averages. We analyze in this investigation in particular the scaling
behavior $\beta = d \log \lambda / d \log E $ of the principal
Lyapunov exponent with the energy of the
lattice system. A large body of recent literature claims 
a linear scaling relationship ($\beta = 1$) between
the principal Lyapunov exponent and the average energy per
lattice plaquette 
for the continuum limit of the lattice Yang-Mills equations.
We question this result by providing rigorous upper bounds
on the Lyapunov exponent for all energies, hence giving
a non-positive exponent, $\beta \leq  0$, asymptotically for high energies,
and we give plausible arguments
for a  scaling exponent 
close to  $\beta \sim 1/4$ for low energies. 
We argue that the region of low 
energy is the region which comes closest to
what could be termed a ``continuum limit'' for the
classical lattice system.

\end{abstract}

\newpage


\setcounter{equation}{0}

The classical Yang-Mills equations for non-Abelian gauge fields provide an
interesting class of dynamical systems for which non-linear
 self-coupling terms
open up the possibility of chaotic behavior even in toy-models where the
gauge fields are spatially homogeneous \cite{MatinyanSavvidy81} - 
\cite{Bermanetal94}. However, a
qualitative important new dynamical feature
comes into play when one considers the spatially inhomogeneous classical
Yang-Mills equations: There is a never ending cascading of the dynamical
degrees of freedom towards the ultraviolet, generated by the time evolution
of the Yang-Mills equations as we shall discuss below.
In spite of this ``ultraviolet catastrophe'', we note that the solutions are
well behaved, in the sense 
that there are
no ``finite time blow up of 
singularities''. For example, the class of
classical Yang-Mills equations in 3+1 dimensions is known to have
solutions in suitable Sobolev spaces, such that their norms do not blow up to
infinity in finite time \cite{Ea,Kl}.
Without an ultraviolet cut-off of the Yang-Mills field equations, 
it is however not possible
- even when the fields are confined to a finite volume -
to  define equilibrium statistical mechanics,
i.e. make reasonable sense out of infinite-time 
time averages of observables and
correlation functions. The reason for this is that although the system
has finite space volume, the phase space volume is still infinite.
As is well-known in the case of electromagnetism,
a reasonable definition of
equilibrium leads to an ultraviolet catastrophe as seen in the
Rayleigh-Jeans law for energy (per frequency) density. 
The arguments in the electromagnetic case are purely thermodynamic
and do not involve the classical evolution of the fields. In fact,
since the corresponding equations are linear the problems would not show up
dynamically unless
one couples the fields to charged particles.
In contrast, the Yang-Mills equations for non-Abelian gauge fields
include non-linear self-coupling terms which open up the
possibility for a chaotic behavior in the classical evolution
and in the non-homogeneous case 
 it leads to the infinite cascade of energy
from the long wavelength modes towards the ultraviolet.
This
is a natural dynamical interpretation of the ultraviolet catastrophe
if one assumes
that on average a trajectory
should simulate
a micro canonical ensemble where, in an equilibrium situation, all modes
have the same energy on average. But
due to the infinite phase space  volume, this  implies that
this average energy per mode 
 has to be zero (!), whence the cascade.
We would like to add that this picture is well supported by
the articles
of Furusawa \cite{Furusawa} and Wellner \cite{Wellner}.

This tendency of the mode frequencies cascading
towards the ultraviolet will
completely dominate the qualitative behavior of the classical Yang-Mills
equations, and the ``ultraviolet catastrophe'' has for some time been
emphasized by us (H.B.N. and S.E.R., 
cf. e.g. discussion in \cite{RughYMlic94})
as a major obstacle to simulate  
the classical continuum Yang-Mills fields in a numerical experiment
over a long time span. (This obstacle has in our opinion received
insufficient 
attention in the various studies 
attempting at discussing and modeling
chaotic properties of spatially inhomogeneous classical Yang-Mills fields).
There is no mechanism, within the classical equations, which prevents this
never ending cascading of the modes towards the ultraviolet.
Nature needs $\hbar$, the Planck constant, as an ultraviolet regulator.
Indeed, both Abelian and non-Abelian gauge fields are implemented as quantum 
theories in Nature.

Here, however, we shall 
discuss the possibility of using a lattice cutoff in a purely
classical treatment to regularize the equations and giving sensible
results
in the limit when the lattice spacing $a$ tends to zero.
 The model
we consider is the Hamiltonian lattice formulation first introduced
by Kogut and Susskind \cite{KogutSusskind} (in a quantum context).
There is still a cascading of modes towards the ultraviolet, i.e. towards
the lattice cut-off, and this ultraviolet cascade will still dominate the
dynamical evolution of smooth initial field configurations.
However, in this Hamiltonian formulation on a large but finite lattice,
the phase space is compact  for any given energy and thus
the system  can reach an equilibrium state among the modes 
(a `thermodynamic equilibrium').
The lattice regularization of the theory opens up for defining 
dynamic and thermodynamic properties,
which are not defined in the classical Yang-Mills field
theory without regularization. It could e.g. be
ergodic (modulo constraints) with respect to the Liouville measure,
in which case it makes 
sense to talk about its micro-canonical distribution
and approximating this by looking at 
`typical' classical trajectories. The fundamental
assumption of thermodynamics asserts that 
on average the two 
approaches give the same result if we have a large system, 
and we may then naturally introduce  correlation functions and 
possibly a correlation length $\xi$ (measured in lattice units)
of the system. One hopes to
define a continuum theory if, 
by judicious choice of the parameters in the system,
one obtains a
physical correlation length in the limit when the lattice
constant goes to zero, i.e.
\begin{equation} \label{finitecorr}
\xi(a, E(a),...) \times a  \; \; \r
\; \; \ell  \neq 0 \; \; , \; \; \mbox{as} \ a\r 0.
\end{equation}
In equation (\ref{finitecorr}) the 
correlation length $\xi$ in the
lattice system is a function of lattice model parameters such as 
lattice spacing $a$, average energy density $E(a)$, etc.
Condition (\ref{finitecorr}) implies
that the correlation length diverges when measured in lattice units
and only if this is the case do
we expect the lattice
system to lose its memory of the underlying lattice structure.
Condition (\ref{finitecorr}) is, of course, just a necessary,
not sufficient condition for the lattice system to approach the
continuum theory. 
For example, the limiting system might still carry
the  symmetries of the lattice. We shall return in more detail
to the important issue of what 
we could mean by a ``continuum limit'' of a classical
lattice regularized non-Abelian gauge field theory.

The Kogut-Susskind Hamiltonian is obtained from the
Wilson action,
approximating the Yang-Mills equations, by keeping a lattice grid
in space while letting time become continuous.
For a derivation
we refer to \cite{KogutSusskind} and  also to \cite{Muller} from
which we adapt our notation.
We will restrict attention to the lattice gauge theory in
3+1 dimensions based on the gauge
group $SU(2)$. One considers a finite size
3 dimensional discretized box, i.e. 
a finite lattice having $N^{3}$ points
where nearest neighbor 
points are separated
by a distance $a>0$.
The phase space corresponding to this
is then a fibered space where the tangent manifold of
the Lie-group $SU(2)$
is assigned to each of the links, $i \in \Lambda$,
connecting nearest neighbor lattice points.
More precisely, we have to each
$i \in \Lambda$ associated a link variable
$U_i \in SU(2)$ as well as its canonical momentum
$P_i \in T_{U_i} SU(2)$. A point in the entire phase space
 will be denoted
\begin{equation}
\PPsi = \{U_i, P_i\}_{i \in \Lambda} \ \in \ 
M = \prod_\Lambda T \; SU(2)\ .
\end{equation}
The resulting Kogut-Susskind Hamiltonian
generating the time evolution of the orbit $\PPsi (t)$ 
can be written in the following way,  where for simplicity
we omit the coupling constant factor $2/g^2$ which anyway is arbitrary in
a classical theory~:
\beq \label{HamiltonKS}
  H(a, \PPsi^{(a)}) =
    \frac{1}{a}
       \sum_{i \in \Lambda} \half \tr (P_i P_i^\dagger) +
    \frac{1}{a} 
   \sum_\plaq(1  - \half \tr U_{\plaq})  \ .
 \label{eq:kogut}
\eeq
Here the last sum is over elementary plaquettes
bounded by 4 links and $U_{\plaq}$ denotes the
path-ordered product of the 4 gauge elements along the
boundary of the plaquette ${\plaq}$.
The last term, the potential term, is automatically bounded and 
for a given finite total energy the same is the case for 
the first term, the kinetic term. Thus the phase space
corresponding to a given energy-surface is compact.

The compactness of the phase space implies that the spectrum of
Lyapunov exponents (which we overall will assume to be well defined 
quantities for the lattice system) is independent of the
 choice of norm on the
space of field configurations. The measures 
of distances between
lattice field configurations $\PPsi (t)$ and $\tilde{\PPsi} (t)$  
which are employed in
\cite{Muller} - \cite{BiroetalBook} are semi-positive and, typically,
 of the form (cf. e.g. \cite{MullerColor})
\begin{equation} \label{MullerDistance}
{\cal D} (\PPsi, \tilde{\PPsi}) \sim \frac{1}{N^3} \;
 ( \; \sum_{i \in \Lambda} |tr(P_i P_i^{\dagger}) -
tr(\tilde{P}_i \tilde{P}_i^{\dagger}) | \; + \;
\sum_{\Box} | tr U_{\Box} - tr \tilde{U}_{\Box} | \; ) 
\end{equation}
which in the limit as the lattice constant $a \rightarrow 0$ 
measures the (average)
local differences in the electric and magnetic field energy,
\begin{equation}
{\cal D} ((E,B), (\tilde{E}, \tilde{B})) \sim \frac{1}{V}
\int d^3 x \; ( | E^2 - \tilde{E}^2| + |B^2 - \tilde{B}^2 |) \; \; .
\end{equation} 
Gauge equivalent field configurations have a vanishing distance since
the distance measure (\ref{MullerDistance}) is gauge invariant.\footnote{As
regards gauge invariance in the resulting measures
of chaos, we note that since
``space-time'' is not involved in the gauge transformations, we expect 
Lyapunov exponents and correlation lengths to be gauge invariant 
measures of temporal and spatial chaos in the context of Yang-Mills fields,
whereas such gauge invariant measures of chaos are more difficult 
to construct for a typical metric in general relativity \cite{Rugh1994}.} 
As discussed at the end of the appendix such a choice of distance
will not affect the calculation of the largest Lyapunov exponent.

It is easy to see that if $\PPsi^{(a=1)} (t) $ 
solves the Hamiltonian equations for $a=1$
then $\PPsi^{(a)}(t)  = \PPsi^{(a=1)}(t/a)$
 solves the same equations
for general $a$ and this
time-scaling implies the following
scaling of the maximal Lyapunov exponent~:\footnote{In the 
equation (\ref{eq:Lyapscaling1}) we should in fact replace
the small deviation vector  $\delta \PPsi$ by a tangent vector
to the manifold, cf. the appendix.}
\begin{eqnarray} \label{eq:Lyapscaling1}
 \lambda_{max}(a) & \equiv &
   \lim_{t \r \infty} \frac {1}{t} \log (\frac{
   \|  \delta \PPsi^{(a)} (t) \|}{ \| \delta \PPsi^{(a)} (0) \|})
     \nonumber  \\ 
   & =  &  
   \frac{1}{a} \lim_{t/a \r \infty} \frac{1}{(t/a)} \log (\frac{ 
   \|  \delta \PPsi^{(a=1)} (t/a) \|}{\| \delta \PPsi^{(a=1)} 
   (0) \|}) = \frac {\lambda_{max}(a=1)}{a} 
\end{eqnarray}
Equation (\ref{eq:Lyapscaling1}) implies that $a \lambda_{max}(a)$ is 
invariant under rescaling of $a$.
If we assume ergodicity over the energy surface 
(modulo the known constraints) we see that in particular,
$a \lambda_{max}(a)$ is uniquely determined by
the energy
$a  H(a) =  H(a=1)$
or equivalently when lattice size $N^3$ is fixed
it has to be some function of 
the energy per plaquette $a E(a) = E(a=1)$,
\begin{equation} \label{LyapfE}
a \; \lambda_{max} (a, E(a)) = f \; (a \; E(a))
\end{equation}
Such a functional relationship even persists in the thermodynamic
limit, $N \r \infty$, 
for any fixed values of $a$ and $E(a)$.
Thus in order to study the dependence of the maximal Lyapunov
exponent with energy density, it is sufficient to consider
the equations of motion for a fixed value of  the lattice constant
$a$, e.g. $a=1$, 
as a function of energy density
($\propto$ energy/plaquette) and
rescale  the results back afterwards.

Studying the classical dynamics in time rather than the ensemble
distribution makes it possible to
extract dynamical information as well as thermodynamic information.
By measuring
the maximal Lyapunov exponent $\lambda_{max}(a)$ for some fixed
lattice spacing $a$
of the system as a function of the
energy, one may also hope to gain insight into the spatial correlations
of the fields. 
Although a precise relationship has not been established yet,
it seems plausible that when propagation of
information to nearest neighbors occurs at a fixed speed,
a small value of $\lambda_{max}$
should lead to a larger correlation length than a larger value of
 $\lambda_{max}$. The reason is that in order
for the dynamics  to create long
range spatial correlations in the system, information has to propagate
for a long time ($=$ distance) without seriously being
attenuated by the chaotic behavior, the `strength' of which
is reflected in the value of $\lambda_{max}$. 
Clearly, it deserves further investigation to establish 
a precise relationship between $\lambda_{max}$ 
and the spatial correlation length $\xi$ (in lattice units).
Such investigations would be of considerable interest also
in the more general case where the non-Abelian gauge fields are 
given a non-zero mass, with an adjustable mass parameter,
 by coupling to a Higgs-field. 
In our case, which involves massless non-Abelian gauge fields, we
expect that it is in the limit as $\lambda_{max}$ goes to zero
that the spatial correlation length $\xi$ 
(in lattice units) will diverge.
   \footnote{It is a generic
  expectation for a chaotic, non-integrable, spatially extended system
  (with a spatial propagation of disturbances at some
  fixed speed) 
  that increasing the  temporal chaos, as captured e.g. by 
  measuring the maximal Lyapunov exponent $\lambda_{max}$, 
  will be  connected to more spatial chaos, as measured, say, by an inverse
  spatial correlation length $\sim \xi^{-1}$. Cf. e.g.
  discussions in \cite{TBohr,Cross}.
  Note, however, that an integrable,
  spatially extended Hamiltonian system can also have finite spatial 
  correlations due to the initial conditions rather than the dynamics. I.e.,
  an integrable
  Hamiltonian spatially extended system will not be able to
  establish spatial correlations
  in field configurations which initially are spatially de-correlated.}

In fact, a complicated question (which needs
further study) is
what we could mean by a
``continuum limit'' for a classical, lattice regularized 
non-Abelian gauge field theory. 
Whereas the issue of 
extracting 
expectation values of observables in the
``continuum limit'' is standard in interpreting
simulations of quantum field theories in the imaginary time
(Euclidean) formalism,
the issue of extracting ``continuum results'' (rather than lattice
artifacts) is much less well-known in the case of real-time
simulations of classical non-Abelian gauge theories. 
Moreover, it is an important question of interest in physics,
since not many non-perturbative methods are available
to study the time evolution of quantum field theories,
such as Q.C.D., or the weak sector of the Standard Model,
and one almost has to resolve to
study the real time evolution of the classical fields - 
an approach approximately justified 
when probing physical situations (e.g. at high temperatures)
where the quantum fields are expected to behave semi-classically
(see also, e.g. discussions in \cite{Bodekeretal}).

For the study of the time evolution of Yang-Mills fields
which initially are far from 
an equilibrium situation, the (classical)
field modes will exhibit a never ending dynamical cascade towards the 
ultraviolet and after a 
certain transient time, the cut-off
provided by the spatial lattice will prevent the lattice gauge theory
from simulating this cascade.
It is therefore immediately clear that
the lattice regularized, classical fields will not approach a
``continuum limit'' in the sense of simulating the dynamical
behavior of the classical continuum fields in the $t \rightarrow \infty$
limit (we may also say that the continuum classical theory 
does not exist in the infinite time limit).

We have several different forms of lattice artifacts in the lattice 
simulation of real-time dynamical behavior of the continuum classical
 Yang-Mills fields:

(1) Lattice artifacts due to the compactness of the group.
 The magnetic term (the second term) in the Kogut-Susskind
Hamiltonian (\ref{HamiltonKS}) is uniformly bounded,
$ 0 \leq 1 - \frac{1}{2} Tr U_{\Box} \leq 2$, due 
to the $SU(2)$ compactification. Thinking in terms of statistical 
mechanics for
our classical lattice system, we expect that after some time the 
typical field configuration has equally much energy in all modes
of vibration - independent of the frequency\footnote{This situation is
very different in the quantum case. 
Planck's constant $\hbar$ introduces a relation
$E = \hbar \omega$ between the energy of a mode of vibration
and its frequency, 
implying that a mode with a high frequency also has a high energy.
With a given available finite total energy,  modes with
high frequencies will therefore be suppressed. 
Quantum mechanically, we thus
have that at low energy only excitations of the longest
wavelengths appear.} - and the total
amplitude of the classical field, and the energy per lattice
plaquette, is thus small for a fixed low energy. For low energy,
when the average energy per plaquette is small, the 
lattice artifacts due to the compactness of the gauge
group are thus negligible.  

(2) For small energy per plaquette, we thus expect that the dominant 
form for lattice artifact is due to the fact
that an appreciable amount
of the activity (for example the energy) is in the field modes
with wavelengths comparable to the lattice constant $a$.
This short wavelength activity at lattice cut-off scales
is unavoidable in the limit of long time simulation of
an initially smooth field configuration (relative
to the lattice spacing),
or already after a short time if we initially have an irregular
field configuration.

We conclude that for the simulation of gauge fields far from 
equilibrium, we will have
the best ``continuum limit'' if we simulate,
for a short period of time, an initial
smooth ansatz for the fields in the region
of low energy per plaquette.\footnote{We are here imagining a 
situation where the spatial correlations in the monitored
field variables are so large that they
lose memory of the underlying lattice structure (including  the lattice
spacing $a$). In the extreme opposite limit,
one could imagine models in situations with random
fluctuating fields on the scale of the lattice constant,
i.e. with (almost) no spatial correlations from link to link.
If the field variables fluctuate independent of each other
(independent of their neighbors) one could imagine the
model to be invariant (with respect to the monitoring of many
variables) under changes of
the lattice spacing, $a$. Thus, it appears that
lattice cut-off independence of numerical results can not be
a sufficient criterion for the results to report ``continuum physics''.} 

This is if we have no quantum mechanics.
However, Yang-Mills fields exist as quantum 
theories. Thus, the interesting physical definition of 
a ``continuum limit'' of the classical
lattice gauge theory, is to
identify  regions in the parameter space for the classical
lattice gauge theory which, for (shorter or longer) intervals of
time, probe the behavior of the time evolution of semi-classical
initial configurations (with many quanta) of the quantum
theory, for example Q.C.D. (See also e.g. discussion in \cite{Bodekeretal}).

There is by now a large amount of literature which reports investigations
of temporal chaos on the lattice gauge theory measured by 
the spectrum of Lyapunov exponents 
(either the principal Lyapunov exponent or the 
entire spectrum of Lyapunov exponents)
and how Lyapunov exponents depend on the average energy per plaquette, $E$,
 of the lattice field theory.
We refer to a sequence of articles by M\"{u}ller et al  \cite{Muller},
Gong \cite{Gong}, Bir\'{o} et al. \cite{Biro} and a recent book 
by Bir\'{o} et al \cite{BiroetalBook} which present
numerical analyses of the classical $SU(2)$ lattice gauge model.
Their numerical results provide evidence
that the maximal Lyapunov exponent is
a  monotonically
increasing continuous function of the scale free energy/plaquette
with the value zero at zero energy.

M\"uller et al. \cite{Muller} report a particular interesting
interpretation of numerical results for the dynamics on the lattice,
namely that there is
a linear scaling relation between the scale free maximal Lyapunov exponent,
$\lambda_{max}(a=1)$ and the  
energy per plaquette $E(a=1)$.
The possible physical relevance of this result is seen when we
rescale back to a variable lattice spacing $a$ and note that
the observed relationship is in fact a graph of
$a \lambda_{max}(a)$ as a function of $a E(a)$. Thus being
linear, cancellation of a factor $a$ implies that 
$\lambda_{max}(a) = \mbox{const} \times E(a)$ 
 and thus there is
a continuum limit $a\r 0$ either of both sides simultaneously
or of none of them. In the particular case where the energy
per mode ($\propto$ energy per plaquette) is taken to be a
fixed temperature $T$ (cf. \cite{Ambjorn}), one deduces that the maximal
Lyapunov exponent has a continuum limit in real time.

We shall here, however, note that the 
linear relationship found between $\lambda_{max}$
and $E$ is based on a graph of their interdependence (indicating
that such a scaling relation exists)  in the approximate 
scaling region 
(for $a=1/2$) of $E$ between 
$0.6$ and $4$,
where the maximal Lyapunov exponent grows
from $0.2$ to $1.3$
 (cf. figure 1)~:
\beq 
   \lambda_{max} \approx 0.32 \; E \ .
   \label{eq:linear}
\eeq

We shall argue 
that the apparent linear scaling relation
is a transient phenomena residing in a
region extending at most a decade
 between two scaling regions, namely 
for small energies 
where the Lyapunov exponent scales with an
exponent which could be close to $1/4$ and a high energy region where
the scaling exponent is at most zero:
\begin{equation}
\beta = \frac{d \log \lambda}{d \log E} =
\left\{  \begin{array}{ll}
\sim 1/4 & \mbox{for $E \r 0$}  \\
\leq 0 & \mbox{for $E \r \infty$}
\end{array}
\right.
\label{eq:our}
\end{equation}
In fact, we
believe that the apparent linear
scaling region, 
equation (\ref{eq:linear}),
is a complete  artifact of the 
$SU(2)$ compactification on a lattice.
We argue by giving
a newly obtained \cite{hhrugh}  rigorous upper
bound for the maximal Lyapunov exponent, as well as 
plausible lower bounds obtained from general scaling arguments 
\cite{RughYMlic94} of the
continuum classical Yang-Mills equations in accordance with simulations on
homogeneous models \cite{Chirikov} and consistent with the figures in
\cite{Muller} and \cite{Biro}.
Thus the existence of the two scaling regions, equation
(\ref{eq:our}), is established 
by a combination of numerical evidence and analytical arguments.
In figure 1 (a) we have plotted the numerical results
reported in M\"{u}ller et al \cite{Muller} which exhibits
a systematic deviation from a straight line.
 A recent numerical investigation by Krasnitz 
(cf. \cite{Krasnitz}, figure 3) has 
independently confirmed such a systematic deviation.
According to Krasnitz
statistically significant deviations from the straight line is as large as
10 \%. 
In figure 1 (b) we have made a log-log plot of the results obtained
by M\"{u}ller et al \cite{Muller}.

Regarding 
 the limit for high energies per plaquette a rigorous result
\cite{hhrugh} (a sketch is shown in Appendix A)
shows that the $SU(2)$ scale free ($a=1$)
 lattice Hamiltonian in $d$
spatial dimensions
 has an upper bound
for
the maximal Lyapunov exponent
\begin{equation} \label{laB}
 \laB = \sqrt{(d - 1) (4+\sqrt{17} )}
\end{equation}
which for $d=3$  becomes $\laB = 4.03..$. 
This result  is arrived at by constructing
an appropriate  norm on the phase
space 
and showing that the time derivative of this norm can be bounded
by a constant times 
 the norm itself, hence giving us an upper bound as to how 
exponentially fast the
norm can grow in time. 
The upper bound (\ref{laB}) shows that a linear scaling
region, i.e. a constant $\beta = d \log \lambda / d \log E$, can 
not extend further than around $E \sim 10$
on the
figure 1. Beyond that point the maximal Lyapunov
exponent either saturates and scales with energy with an exponent which
approaches zero or it may even decrease
over a region of high energies, yielding a negative
 exponent\footnote{
             Some very heuristic arguments yields
            $\beta = -1/2$ asymptotically.}.
It should be noted that the upper bound 
(\ref{laB}) is independent of the lattice size and the energy
(but scales with $1/a$).

In the opposite limit, the ``continuum limit'', 
where the average energy per plaquette $ E \rightarrow 0$,
the finite size of the lattice makes
it much more  difficult to 
analyze the behavior of the gauge fields and the principal Lyapunov
exponent on the lattice.
However, we shall argue that 
the scaling exponent of the Lyapunov exponent more likely will be closer to
$\sim 1/4$ 
than the scaling exponent $\sim 1$ observed in the
intermediate energy region $1/2 \; \simleq E \; \simleq 4$. 

In figure 1 (b) we have made a log-log plot of the
results obtained by M\"{u}ller et al. \cite{Muller}. It appears
that for $E \sim 1/2$ there is a cross-over to
another scaling region. Although the data points in this
region are determined with some numerical
uncertainty \cite{Mullerpriv}
we note that they are consistent with a
 scaling with exponent $\sim 1/4$. Moreover,
it is a numerically established 
fact that the homogeneous Yang-Mills equations have a non-zero
Lyapunov exponent which by elementary scaling arguments
scales with the fourth root of the energy density and has the
approximate form \cite{Chirikov}:
\begin{equation} \label{ChirikovFormula}
\lambda_{\rm max} \approx  0.38 \; E^{1/4} \ .
\end{equation}
On the lattice  the above scaling relation is valid for spatially homogeneous
fields, i.e. the maximal Lyapunov exponent scales with the fourth root of the
energy per plaquette.
By continuity, fields
which are almost homogeneous\footnote{Note that such field
configurations are (ungeneric) examples of field configurations
which exhibit correlation lengths much larger than the lattice spacing.} 
on the lattice
will, in their transient, initial dynamical behavior,
exhibit a scaling exponent close to $1/4$.
In fact, the same  scaling exponent\footnote{In the lattice simulations 
the energy of a given plaquette is a fluctuating quantity
and since the
function $E^{1/4}$ is convex one would expect the
coefficient of proportionality to be somewhat smaller than in
formula (\ref{ChirikovFormula}).
Numerically the coefficient 
turns out to be around half of the above value,
cf. figure 1 (a-b).} 
would also hold for
the inhomogeneous Yang-Mills equations \cite{RughYMlic94} had
the fields been smooth relative to the lattice scale,
so derivatives, $\partial_\mu$, are well approximated by their
lattice equivalent and
we are allowed to scale lengths
as well. This is seen from scaling arguments for the continuum Yang-Mills
equations~:
\[ D_\mu F^{\mu \nu} = 0 \ \ , \ F^{\mu \nu} = 
      \partial^\mu A^\nu - \partial^\nu A^\mu - ig[A^\mu,A^\nu]
   \ \ , \ D_\mu = \partial_\mu - i g [A_\mu, \; \; ] \]
Not taking boundary conditions into account,
these equations are invariant when $\partial_\mu$ and $A_\mu$ are
scaled with the same factor $\alpha$. That is,
if $A(x,t)$ is
a solution to the equations, then 
$\frac{1}{\alpha} A(\alpha x, \alpha t)$ is also a solution. 
The energy density $E$,
which is quadratic in the Yang-Mills
field curvature tensor $F^{\mu \nu}$,
then scales with $\alpha^4$.
The same arguments as leading to equation (\ref{eq:Lyapscaling1})
then show that if we perform 
a measurement of the maximal
Lyapunov exponent $\lambda_{max}$ over a time short enough  for
the solutions to stay smooth, then $\lambda_{max}$ scales with
$E^{1/4}$, as was also observed by 
M\"{u}ller et al. 
p. 3389 in \cite{Muller}. These scaling arguments
do not carry over to infinite time averages since solutions on
the lattice tend to be irregular.
We believe, however,  that the  cross-over observed in figure 1(b)
is a feature of the Kogut-Susskind model
and not just a finite size
effect, in particular since there is evidence \cite{Ambjornpriv}
that the  correlation
length for energies $E \sim 1/2$ is of the order of 
a few lattice units only. 
As the numerical simulations of M\"uller et al.
are performed  on lattices $\sim 20^3$
and since the calculation of Lyapunov exponents is a 
local calculation
(when considering infinitesimal deviations),  
a finite size effect should show up
for energies somewhat below $E \sim 1/2$.

The expectations, arrived at here, 
are supported by 
an even more striking result, which is
the numerical simulations of 
Bir\'{o} et al 
(\cite{Biro}, figure 12)
for the Kogut-Susskind lattice model with a $U(1)$ group showing
(see figure 2) a steep increase of the maximal Lyapunov exponent
with energy/plaquette in the 
interval $1 \; \simleq \; E \; \simleq \; 4$. 
The continuum
theory here
corresponds to the classical electromagnetic fields which have no
self-interaction and thus the Lyapunov exponent in this limit
should vanish.
 The discrepancies in this case were in \cite{Biro}
 attributed to a combined effect of the discreteness of the lattice and
the compactness of the gauge group $U(1)$  and were not
connected with finite size effects. This suggests strongly
that in the case of $SU(2)$ a similar cross-over around
$E \sim 1/2$ 
 should have the same origin and likewise, not be seen
as a finite size effect.
In particular,
it makes it difficult to believe that we in the case of
numerical studies of a $SU(2)$ Kogut-Susskind Hamiltonian system can base
continuum physics on results from simulations in the same interval of
energies where the $U(1)$ simulations fail to display continuum
physics.
On the contrary, we suspect that 
what could reasonably be called ``continuum physics'' has to be extracted from
investigations of the Kogut-Susskind lattice simulations for energies 
per plaquette which are
at least smaller than $ E \sim 1/2$. 
In conclusion, we believe that there is indeed
a cross-over around $E \sim 1/2$  
but we
do not attribute this to finite size effects\footnote{It would be valuable
to study the spatial correlation lengths on the lattice system for
energies per plaquette in the interval $0 \leq E \simleq \;1$, i.e.
for an interval
of energies which includes the apparent cross-over at
$E \sim 1/2$. }
which would invalidate the physical relevance
of the lower region. Rather, we believe and conjecture
that this phenomena will persist even for infinitely big lattices
still showing a cross-over around $E \sim 1/2$,
below which we approach ``continuum physics''  
and above which
discreteness of the lattice and compactness of the gauge group
are no longer negligible.

There is quite a simple intuitive explanation for the
kind of transition taking place for energies in the region
above  $1/2$ for the $SU(2)$ lattice gauge model and
for the saturation of the Lyapunov exponent in the regime for
high energy per plaquette: The Hamiltonian (\ref{eq:kogut}) consists of
two terms, the first one is a kinetic energy term (electric term) which is
a quadratic form in the momenta, the second is a potential term (magnetic
term)
which is uniformly bounded
 $ 0 \leq 1 - \half  Tr U_{\Box} \leq 2$, due to the $SU(2)$ compactification.
For small energies per plaquette, $E \ll 1$, 
both terms will be of comparable size.
In fact, as energy increases from zero, the potential energy
increases too, but as long as the energy/plaquette is much smaller
than 1, the gauge holonomy $U_{\Box}$ calculated around a Wilson loop
is close to the identity so that field curvatures are small.
The dynamics is then essentially confined to 
the Lie algebra, i.e.
the tangent plane of the identity element in $SU(2)$.
  However, since the Lie-group itself is curved, higher order
non-linear terms become important as the energy per plaquette
increases further.
Although this does not imply 'more chaos', it does suggest
a steeper increase in the Lyapunov exponent. As energy increases
even more, $E \gg 1$,  the $U_{\Box}$ will experience the finiteness of the
group $SU(2)$ and energy will only be pumped into the
kinetic term. We note that if only the kinetic term had been present
in the Kogut-Susskind Hamiltonian (\ref{HamiltonKS}), 
the system would be integrable. 
Since  the potential term 
only provides a uniformly bounded
perturbation of the dynamics, it is reasonable to expect
(and shown in the appendix)  that the
spectrum of Lyapunov exponents will saturate as the energy per plaquette
increases for the Kogut-Susskind model. The chaos generated by the non-linear
potential energy term
does not increase, only the energy in the kinetic energy
(the electric fields) increases.

Finally, we will give a few comments on possible 
implications of the scaling relation
(\ref{eq:our}) proposed in this investigation. Since by intrinsic 
scaling arguments (cf. equation (\ref{LyapfE})) 
one has a functional relation between
$a\lambda_{max}(a)$ and $aE(a)$, a $1/4$ scaling for small energies
would imply that
$ \lambda(a) \propto a^{-3/4} E(a)^{1/4} $, 
in which case one cannot
achieve a continuum limit simultaneously for the maximal
Lyapunov exponent and the temperature (assuming that
it is proportional to the average energy per plaquette
 $E(a)$, cf. \cite{Ambjorn}), in particular,
the former would be divergent if the temperature is kept fixed.
There is no particular contradiction in this statement, however,
as there is, a priori, no 
 need for having a finite
Lyapunov exponent in the continuum limit. The erratic and fluctuating behavior
of the fields 
one expects in time as well as in space 
(for numerical evidence, cf. also \cite{Wellner}) on very small
scales indicates  that a Lyapunov exponent will not
be well defined. Clearly, this question deserves further investigation and
we do believe that it is of considerable interest to understand 
in which sense we can or cannot have a ``continuum limit'' of a
spectrum of Lyapunov exponents for a (non-dissipative) field theory
like the $SU(2)$ Yang-Mills theory when we study, say, the 
dynamical behavior in a given finite
(physical) volume $\sim L^3 \sim (N a)^3$ with
periodic boundary conditions in the limit of long time 
$t \rightarrow \infty$. 
A question of equal interest which may
be related to the question above is to understand 
the (physical) spatial correlations in 
the system.
Studies of (semi)classical dynamically generated
temporal and spatial chaos may be important for the understanding
of randomness in space and time of field configurations in Q.C.D.
(see also discussions in e.g. \cite{BiroetalBook,RughYMlic94}).
The present investigation 
has been devoted to a discussion of 
a possible ``continuum limit''
of (chaotic aspects of)
 purely {\em classical} Yang-Mills fields regularized on a spatial lattice,
whereas Yang-Mills fields in Nature are implemented
as quantum fields, regularized by $\hbar$
(supplemented by a regularization/renormalization prescription).
A connection
between (semi)classical aspects of the real quantum theory and 
the behavior of the fields in the lattice
regularized classical field theory would be most interesting
to establish on a more rigorous basis.  

{\em Acknowledgement.} We thank Alexander Krasnitz and
Salman Habib
for several discussions on the subject and
we thank Berndt M\"{u}ller for sending us
numerical data from Ref. \cite{Muller}. S.E.R. would like to thank
the U.S. Department of Energy and Wojciech H. Zurek at the Los Alamos National
Laboratory for support.

\subsubsection*{Appendix}

We shall briefly sketch the proof that for an $SU(2)$
lattice Hamiltonian model (\ref{eq:kogut}) with a fixed
lattice constant 
($a=1$), the maximal Lyapunov exponent is
uniformly bounded from above,
independent of the energy and
the lattice size. We refer to \cite{hhrugh} for
details in the proof.\\

First we shall illustrate
the main trick which is quite trivial: 
Consider a harmonic oscillator
 whose evolution is governed
by the equations $\dot{x} = y$ and $\dot{y}=-\omega^2 x$. This
system is integrable and thus it has vanishing Lyapunov exponents.
Now,
 a way to show this is to consider the norm on the tangent space
(which here is the same as the space itself)
$\|(x,y)\|^2 =
\omega^2 x^2 + y^2$ which for $\omega$ non-zero
is equivalent
to any other norm on $R^2$. The time-derivative of this norm
is given by $\frac{d}{dt} 
\|(x,y)\|^2 = 2 \omega^2 x y + 2 (- \omega^2 x) y = 0$ and hence the norm
is invariant under the flow. It follows that all Lyapunov
exponents vanish.

In the case of $SU(2)$ lattice gauge dynamics, the 
momentum $P$ 
will play the role of $\omega$ above and thus be 
part of an integrable dynamics 
(physically one may think of a rotor on
a sphere). From this integrable part, we shall
construct a `good' metric for which norms of vectors in
the tangent space would be constant in time, had it not been for
the non-linear coupling through the plaquettes.
Taking the time derivative of the
norm, we get a variety of terms which, due to 
the compactness of $SU(2)$ and hence uniform boundedness
of the couplings, can
be bounded by our `good' metric itself ! This kind of bootstrap
argument will give us a uniform
upper bound as to how fast the norm can grow, and
from this we deduce rigorous bounds on the maximal
Lyapunov exponent.\\

The Lie group $SU(2)$ 
will be considered in the quaternion representation,
which is a four dimensional embedding using Pauli matrices
(not explicitly shown here) for which  one writes~:
\[ U = \left( \begin{array}{cc}
           u_0 + i u_3, & iu_1+u_2 \\
           iu_1 - u_2, & u_0 - i u_3 \\
           \end{array}
        \right)  \equiv (u_0,u_1,u_2,u_3)  \]
and define  the usual Euclidean
inner product in $R^4$ by~:
\[ \lUVr \equiv u_0 v_0 + u_1 v_1 + u_2 v_2 + u_3 v_3
       = \half \tr (U V^\dagger) \ .\]
This yields a representation of
the 6 dimensional  tangent manifold $T\; SU(2)$, such  that
a base point
$U \in SU(2)$ and a tangent vector
$P \in T_U SU(2) $ 
are given
coordinates $(U,P) \in R^4 \oplus R^4$, subjected to the constraints~:
\[ \lUUr \equiv 1 \ \ \mbox{and} \ \ \ \lPUr \equiv 0 \ \ .\]
The entire phase space is denoted
\[ M = \prod_{\Lambda} T\; SU(2) \ .\]
Using the above notation, the Hamiltonian
(\ref{eq:kogut}) can be written as follows~:
\[ H = \sum_{i \in \Lambda}
 \half \lr{P_i}{P_i} + \sum_\plaq (1 - \half \tr U_\plaq)\ , \]
where the lattice constant is set to $a=1$.
The variation of $H$ with respect to $U_i$ gives us
 $\delta H = \lr{\delta U_i}{V_i}$ 
and thus a force acting upon $U_i$~:
\[ V_i = \sum_{(jkl)} U_j U_k U^\dagger_l \ .\]
The sum above extends over all neighboring links, such
that $(ijkl)$ forms a plaquette. Hence
if the dimension of the space is $d$, $V_i$ depends in total of
$2 (d-1) (4 - 1)= 6(d-1)$
neighboring links but not on $U_i$ itself.

Under the constraints on $U$ and $P$ given above, the
Hamiltonian gives rise to a flow 
$\phi^t : M \r M$, $t\in R$ determined by the following
differential equations (see e.g. \cite{BiroetalBook})~:
\begin{equation} \label{eqsappendix}
  \begin{array}{lcl}
  \dot{U_i}  &=&  P_i \ \ , \\
  \dot{P_i}  &=&  V_i - U_i \lUiVir - U_i \lPiPir \ \ .
  \end{array}
\end{equation}
Consider 
a small deviation 
$\delta \PPsi$ of a trajectory
$$\PPsi + \delta \PPsi =\{U_i+\delta U_i,P_i+\delta P_i\}_{i \in \Lambda} 
\in M$$ and linearize the above equation 
(\ref{eqsappendix}) in $\dP$ to
obtain the evolution now for
 a tangent vector which we again, by slight abuse of notation, denote
$\dP = \{x_i,y_i\}_{i\in \Lambda} \in T_\PPsi M$ 
\begin{eqnarray*} \label{evolutiontangent}
 \dot{x_i} &=& y_i \ \ ,\\
 \dot{y_i} &=& v_i - U_i \lUivir - x_i \lUiVir \\
           &-& U_i \lVixir - x_i \lPiPir - 2U_i \lPiyir \ . 
\end{eqnarray*}
A 'good' norm is now given
by the following Riemann metric~:
\begin{equation} \label{good}
G(\dP,\dP) =\half \sum_{i \in \Lambda}
       \left( \lxixir (C + \lPiPir) + \lyiyir
           - \lxiPir^2  - \lyiUir^2 \right) \ . 
\end{equation}
with an optimal choice of
 $C = 2 \sqrt{17} (d-1)$. 
For any fixed energy, $H$, $\lPiPir$ is bounded from
above by $2 H$, and a small calculation
shows that $G$ is positive definite as a quadratic form
on $TM$. As we have noted earlier, our phase space is
compact for any fixed energy
 and it follows that $\sqrt{G}$ provides
a norm which is 
 equivalent to
any other norm on $TM$.
If $\|\dP\|$ is another norm, compactness implies
the existence of $k_2,k_1>0$ such that~:
\begin{equation} \label{EquivNorms}
k_1 \|\dP\| \leq \sqrt{G(\dP,\dP)} \leq k_2 \|\dP\| \ .
\end{equation}
In particular, the maximal Lyapunov exponent measured with
any two equivalent norms are identical. From (\ref{EquivNorms})
we have~:
\begin{equation}
 {\lim_{t \r \infty}}\  \frac{1}{t}\  \log 
      \frac{\|\dP[t]\|}{\|\dP[0]\|} =
   {\lim_{t \r \infty}}\  \frac{1}{2t}\  \log
      \frac{G ((\dP,\dP)[t])}{G((\dP,\dP)[0])} \ .
\end{equation}
Now, using the constraints above, one verifies that
the derivative of $G$ along the flow
equals~:
\begin{eqnarray*}
  \frac{d}{dt} G(\dP,\dP)
      &=&  \sum_i \lxiyir (C - \lUiVir) + 
            \lxixir \lPiVir \\
      &-& \lxiPir \lVixir + 
           \lviyir - \lviUir \lUiyir \ \ . \label{eq:flow}
\end{eqnarray*}
The $v_i$ entering the equations above is the tangent vector
associated with $V_i$ and hence, it is a sum of
$6(d-1)$ terms of the type
 $x_j U^\dagger_k U^\dagger_l$
involving the tangent vectors $x_j$ of the neighboring 
lattice sites (cf. above).
A tedious but straight-forward calculation shows
that the above time derivative satisfies~:
\[ 
  | \frac{d}{dt} G(\dP,\dP)| \leq 2 \lambda_B G(\dP,\dP) \]
with $\lambda_{B} = \sqrt{d-1}\sqrt{1+\sqrt{1+\frac{1}{16}}}$.
One thus 
has the bound on the growth rate of the norm~:
\[\sqrt{G(\dP[t],\dP[t])} \leq \exp (\lambda_B t) \sqrt{G(\dP[0],\dP[0])} \ .\]
In the case
of $d=3$, we have $\lambda_{B} = 4.03...$
which therefore gives an upper bound to the maximal
Lyapunov exponent, independent of the energy and the lattice size.

The metric we have constructed above is not gauge invariant
and in principle one would like to study the dynamics
and the metric of the associated 
moduli space, i.e. the space  $M$ modulo gauge transformations.
However, in the study of 
upper bounds for Lyapunov exponents, this distinction is
not necessary since gauge transformations are symmetry transformations
of the dynamics and hence the associated Lyapunov exponents
vanish. This is nicely illustrated by Gong in \cite{Gong},
where the complete Lyapunov spectrum is calculated for $2^3$ and
$3^3$ lattices. One third of the Lyapunov exponents vanish due to
the gauge symmetry, expressed by the Gauss law constraint and residual
static gauge transformations. 
A compactness argument again shows that
using a gauge invariant distance measure
(like the one used in \cite{Muller,MullerColor})
which is only semi-positive but
which distinguishes gauge non-equivalent configurations
will yield the same measurements of Lyapunov exponents
as a strictly positive metric.

\newpage

\section*{Figure captions :}
\mbox{}\\[5mm]

\noindent
Figure 1(a) shows for $SU(2)$ a plot of the maximal Lyapunov
exponent as a function of the average energy per plaquette.
The data points (diamonds) are adapted from M\"{u}ller et al. 
\cite{Muller} (see also Bir\'{o} et al. \cite{BiroetalBook}, p. 192)
where results are obtained from 
a simulation on a $20^3$ lattice with 
periodic boundary conditions in all spatial directions. 
The solid line is a linear fit
through the origin and  the dotted line is the function
$\half \times 0.38 E^{1/4}$ which is half of the result 
(cf. formula (\ref{ChirikovFormula})) obtained from
the homogeneous case
(see also figure 1 (b)). \\[5mm]

\noindent Figure 1(b) 
displays a log-log plot of the same data points as in
figure 1(a). On the figure
we also show the (theoretically obtained) upper bound for the maximal
Lyapunov exponent.
  \\[5mm]

\noindent Figure 2(a) shows a plot similar to figure 1(a) for $U(1)$.
The data points are adapted from Bir\'{o} et al. \cite{Biro}.
We note, that the Lyapunov
exponent grows, as a lattice artifact,  in
exactly the same region of energy/plaquette $1 \leq E \leq 4$ 
where the  figure 1(a) (or 1(b)) is reported to exhibit continuum 
physics. \\[5mm]

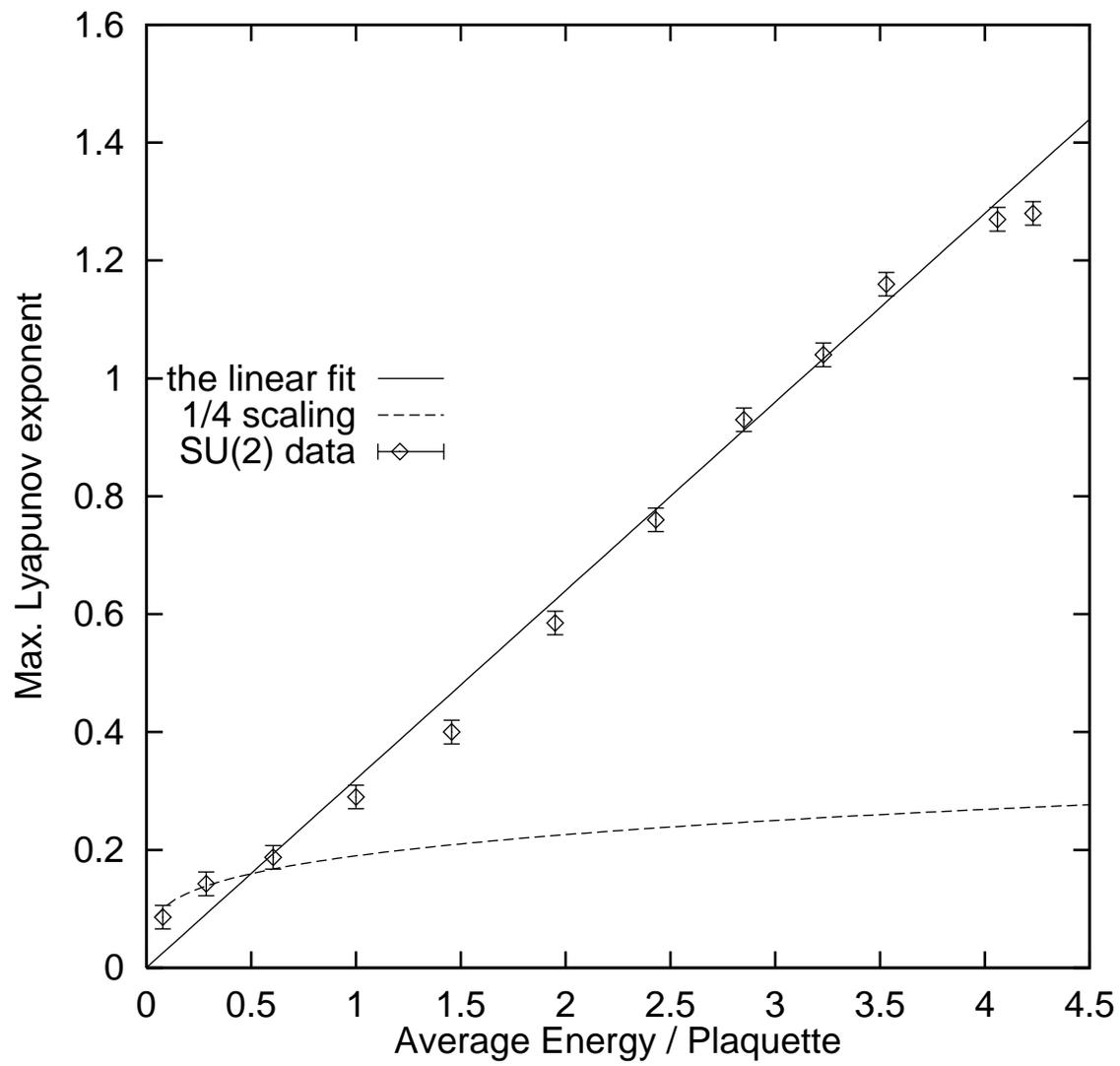
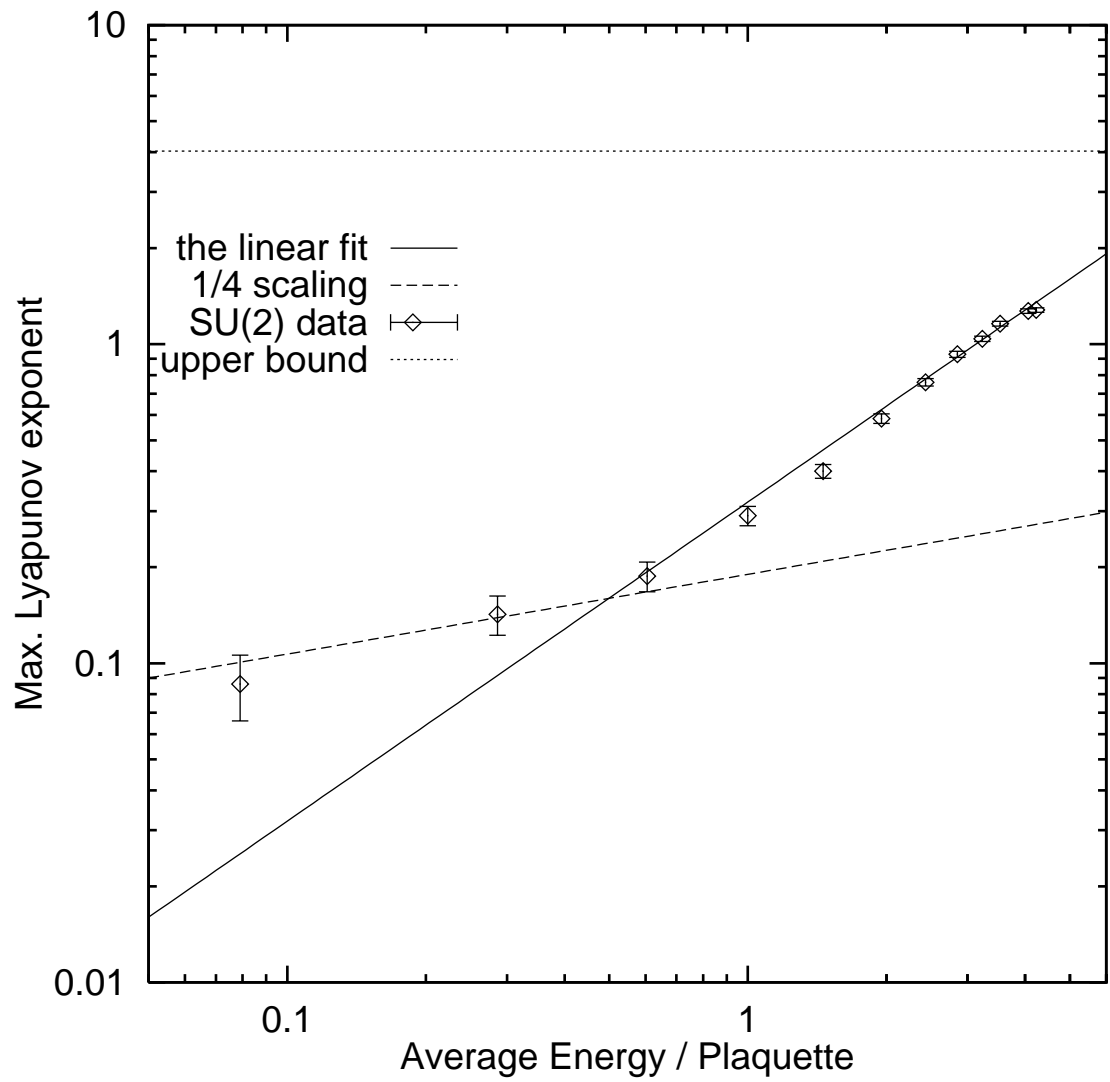
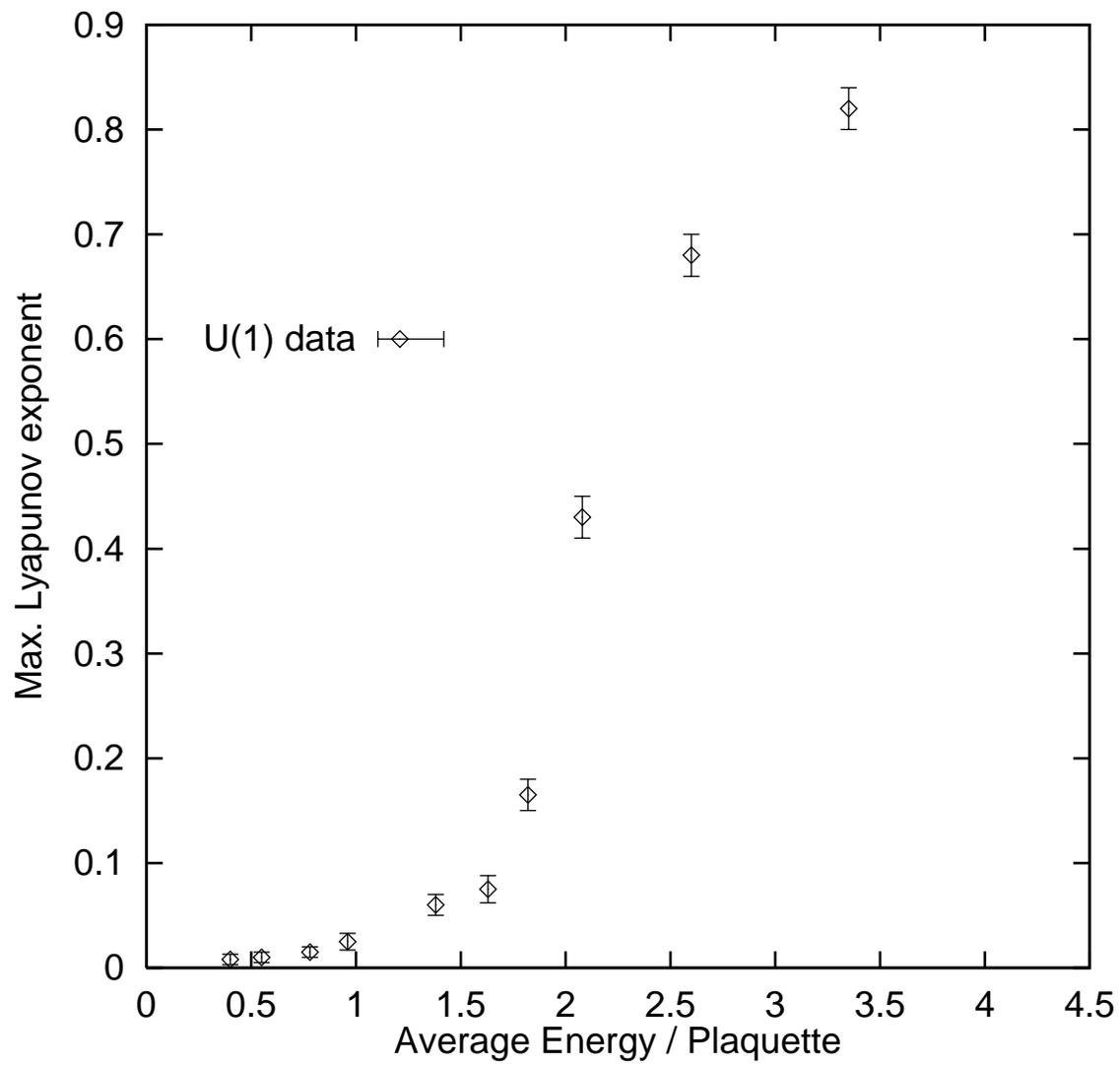
\end{document}